\documentclass[a4paper]{jpconf}
\usepackage{graphicx}
\usepackage{wrapfig}

\begin{document}
\title{The background simulation of experiment for searching of $2\nu2K$ capture in $^{124}$Xe}

\author{V.V.~Kazalov$^{1}$, A.M.~Gangapshev$^{1}$, Yu.M.~Gavrilyuk$^{1}$, V.V.~Kuzminov$^{1}$, S.I.~Panasenko$^{2}$, O.D.~Petrenko$^{2}$, \\
S.S.~Ratkevich$^{2}$,
D.A.~Tekueva$^{1}$
}
\address{$^1$ Baksan Neutrino Observatory INR RAS, Russia}
\address{$^2$ V.N.Karazin Kharkiv National University, Ukraine}

\ead{v.kazalov@gmail.com} 

\begin{abstract}
We describe the Monte Carlo (MC) simulation package of the ``2K-CAPTURE'' setup and discuss the agreement of its output with data. The ``2K-CAPTURE'' MC simulates the energy loss of particles in detector and components of the passive shield and generates the resulting response in working volume large proportional counter (LPC).
The simulation accounts for absorption, reemission,
and scattering of both photons and neutrons and tracks them until they either are absorbed.
The algorithm proceeds with a detailed simulation of the electronics chain. The MC is tuned using data collected with radioactive calibration source
deployed in the internal channel of the installation.
The simulation reproduces the energy response of the detector corresponding to distribution of the generated pointwise clusters of a  charge  of primary ionization in LPC.
\end{abstract}

\section{Introduction}

A unique state is formed in the daughter atom in the case of the capture of two electrons from the $K$-shell.
This state represents a neutral atom with the inflated
shell, exposing two vacancies in the $K$-shell. In order to
detect such a process, we have to keep in mind that for
$Z > 30$, where $K$-fluorescence yields are large, the dominant
decay of double $K$ vacancy states happens through
the sequential emission of two characteristic fluorescence quanta.

The primary contribution (76.7\%)
to the $ECEC$
process in $^{124}$Xe is produced by the capture of two electrons
from the $K$-shell \cite{rf0}. The result of the
$^{124}$Xe($2\nu,2e_K)$ reaction is a neutral $^{124}$Te atom with a
``lifted'' shell, which leaves both $K$-shell vacancies
exposed. The residual excitation of the atomic shell in
daughter isotope $^{124}$Te$^{**}$ relaxes via the emission of
Auger electrons ($e_A$, $e_A$), a single characteristic quantum
and an Auger electron ($K$, $e_A$), or two characteristic
quanta and low-energy Auger electrons ($K$, $K$, $e_A$).
In actual experiments, the almost simultaneous emission
of two characteristic fluorescence quanta produced in the filling of two vacancies offers considerable advantages in terms of detection of such events.

The detection of the relaxation response of atomic
processes after the capture
of atomic electrons in a gas medium offers several significant advantages over liquid
and solid-state detectors.
At the same time, the
impact of the background depends strongly on the
energy resolution, which gives the event detection in a
gas medium another advantage over detection in liquid.
In addition, the interaction of radiation in gas
provides an opportunity to determine the topological
signature of a rare event, 
that  has been demonstrated in
our earlier studies \cite{rf1,rf2}.
In contrast, extremely small
primary ionization tracks typical of interactions in liquid
make it difficult to identify the specific features of
the signal from low-energy primary particles.

The search technique in our experiment is based to the registration triple coincidences of ``shaked'' electrons and two fluorescence photons
produced in the process
of filling of a double $K$-shell vacancy in daughter
atoms.
A proportional counter filled with the gas containing
the studied isotope is used as a detector.
Useful events have a unique feature set in such a detector. Their total signal
comprises three partial pulses with known amplitudes
formed as a result of absorption of two characteristic
photons and a cascade of low-energy Auger
electrons within the working volume. The selection of
events with such features from the entire set of digitized
pulses allows one to raise the effect/background
ratio up to several thousand times. Thus, in our case, the main contribution of the background in the energy region of interest stems from three-point events in the detector.
The main contribution to such events can result from the processes of double ionization of the $K$-shell or the emission of a low-energy gamma-quantum with simultaneous registration of the relaxation products of the daughter atomic shell.

\section{Setup description}
\begin{wrapfigure}[27]{l}{0.5\textwidth} \vspace{0.45pc}
\includegraphics[width=13.0pc,angle=270]{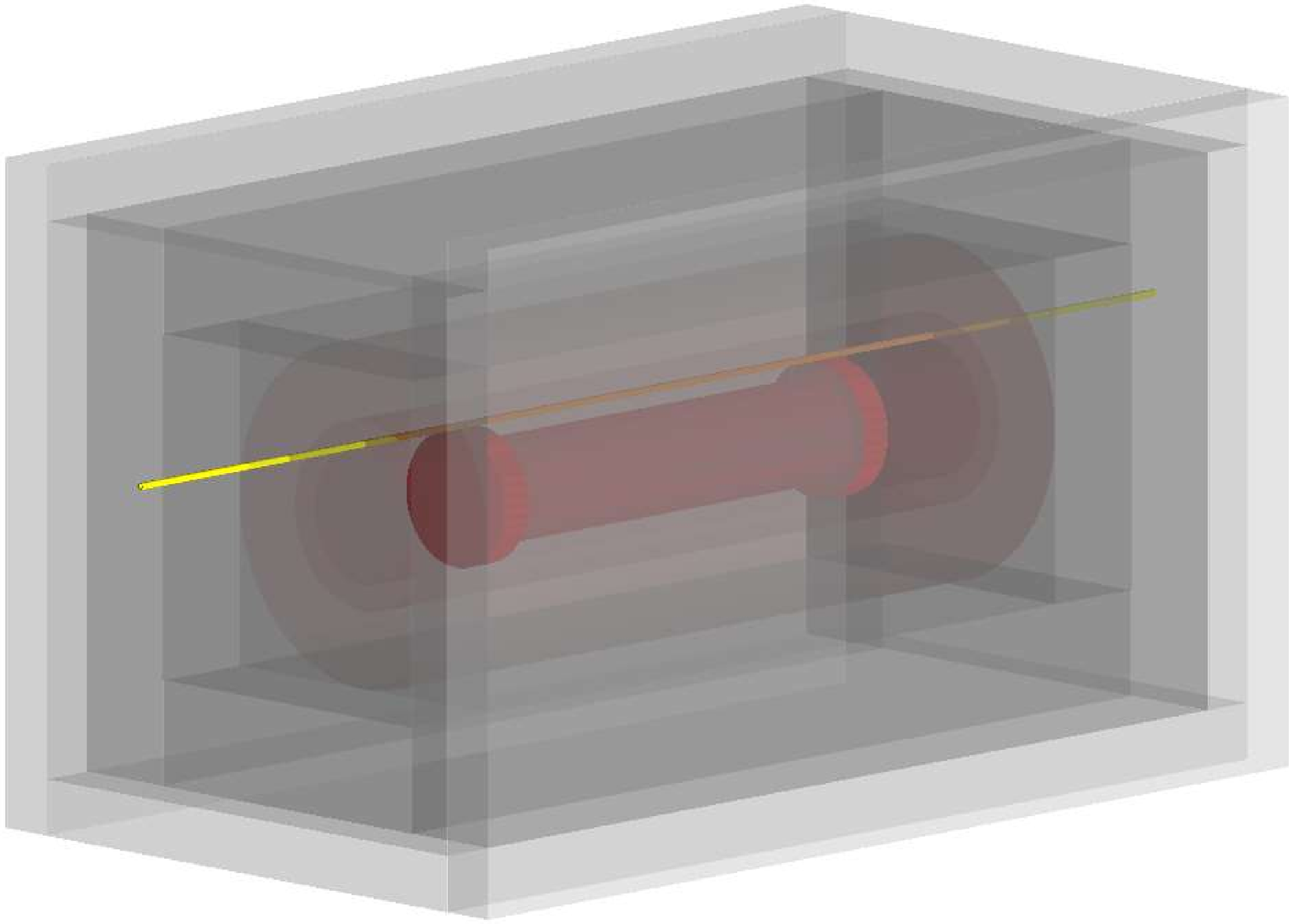}
\caption{\label{pic1} {\small Simulated geometry for the installing ``2$K$-CAPTURE''
 with a calibration channel in Monte-Carlo Geant4 package.}
}
\includegraphics[width=5.0pc,angle=270]{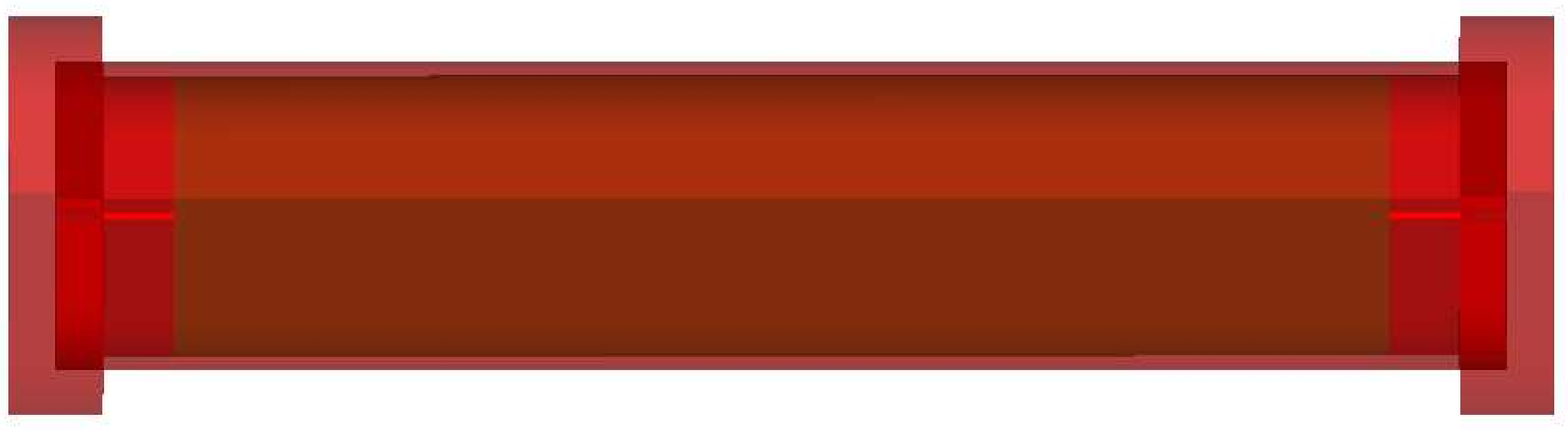} \vspace{0.45pc}
\caption{\label{pic2} {\small Model of large proportional counter. A darker color indicates the operating area of the detector.}
}
\end{wrapfigure}

The model of the experimental setup "2K-CAPTURE" was created with the help of the Monte-Carlo Geant4 package  \cite{r1} for studying the detector response to decays of radionuclides from U/Th families. The same model was used for simulate neutron transport and scattering.
Figure \ref{pic1} shows a cross-section of the
setup as implemented in the simulation.
All the most
relevant geometrical features are included, such as the
multilayer shield with their real shapes and positions. The experimental setup consists of the large proportional counter (LPC) with a casing made of M1-grade copper (the inner surfaces of a casing
and flanges were covered additionally with 1.5 and 3 mm layers of M0k copper  correspondingly)  and  passive  shield.  The shield consists of 18 cm of copper, 15 cm of lead and 8 cm of borated polyethylene.

Additional constructive parts located on the detector flanges, which have high voltage inputs designed to supply high voltage to the anode of the LPC, as well as the pickup of the signal through a high-voltage separating capacitor, were not taken into account in the model.
In the model, the body of the LPC is presented as a 710-mm-long copper cylinder with a working length of 595 mm and an inner diameter of 137 mm, closed on both sides with flanges, Fig.\ref{pic2}
The following structural parts were taken into account in the internal volume of the LPC: a) the anode  wire made from gold-plated tungsten with  a  diameter  of  10 $\mu$m thick is stretched along the cylinder axis; b) the anode  wire  passes through the copper tubes along the edges of the detector, which in turn is inserted into the Teflon insulators.
The gas volume of the detector is represented by two logical volumes: the full volume and the fiducial volume of the detector, which is part of the full volume and is located between the copper tubes in which the anode wire is included. The same volume is a sensitive area of the detector, in Fig.\ref{pic2} it is represented in a darker color.

The LPC is filled with xenon gas to a pressure of 4.8 atm. Xenon has a certain isotopic composition that was taken into account when creating the model. The isotopic composition of gas is presented in Table 1.
\begin{table*}
\centering
\caption{\label{jfonts}Isotope composition of working gas of xenon.}
\begin{tabular}{@{}l*{15}{l}}
\br
{Isotope} & {\small{124}} & {\small{126}} & {\small{128}} & {\small{129}} & {\small{130}} & {\small{131}} & {\small{132}}  & {\small{134}}  & {\small{136}} \\
\mr
Content, \%
&20.311&27.12&33.44&18.812&0.071&0.057&0.026&0.088&0.0806\\
\br
\end{tabular}
\end{table*}

As mentioned above, the LPC is surrounded by multi-layer low-background shield.
For example, the copper shield also has several volumes.
Therefore, to simplify the simulation: \emph{a}) a copper layer around the detector housing located between the opposing flanges; \emph{b}) copper layers at the ends of the detector; \emph{c}) a layer of copper, which surrounds the detector and copper portions at the ends of the detector, has calibrated holes in this volume, and through it a calibration channel made of Teflon.

\section{Simulation results}
The following libraries were used for modeling and calculations:
$G4DecayPhysics$ - for modeling decays of unstable particles indicated in PhysicsList, as well as for all unstable particles that may occur during the simulation; $G4RadioactiveDecayPhysics$ \cite{r2} - allows to simulate radioactive decays of various isotopes, based on ENSDF data \cite{r3}; the $G4EmPenelopePhysics$ library was used to simulate electromagnetic processes, and $G4EmLivermorePhysics$ (for test) was also considered; The $G4HadronPhysicsQGSP\_BIC\_HP$ library was used to \begin{wrapfigure}[24]{l}{0.55\textwidth} \vspace{0.0pc}
\vspace{-0.0pc}
\centering
\includegraphics[width=14.75pc,angle=0]{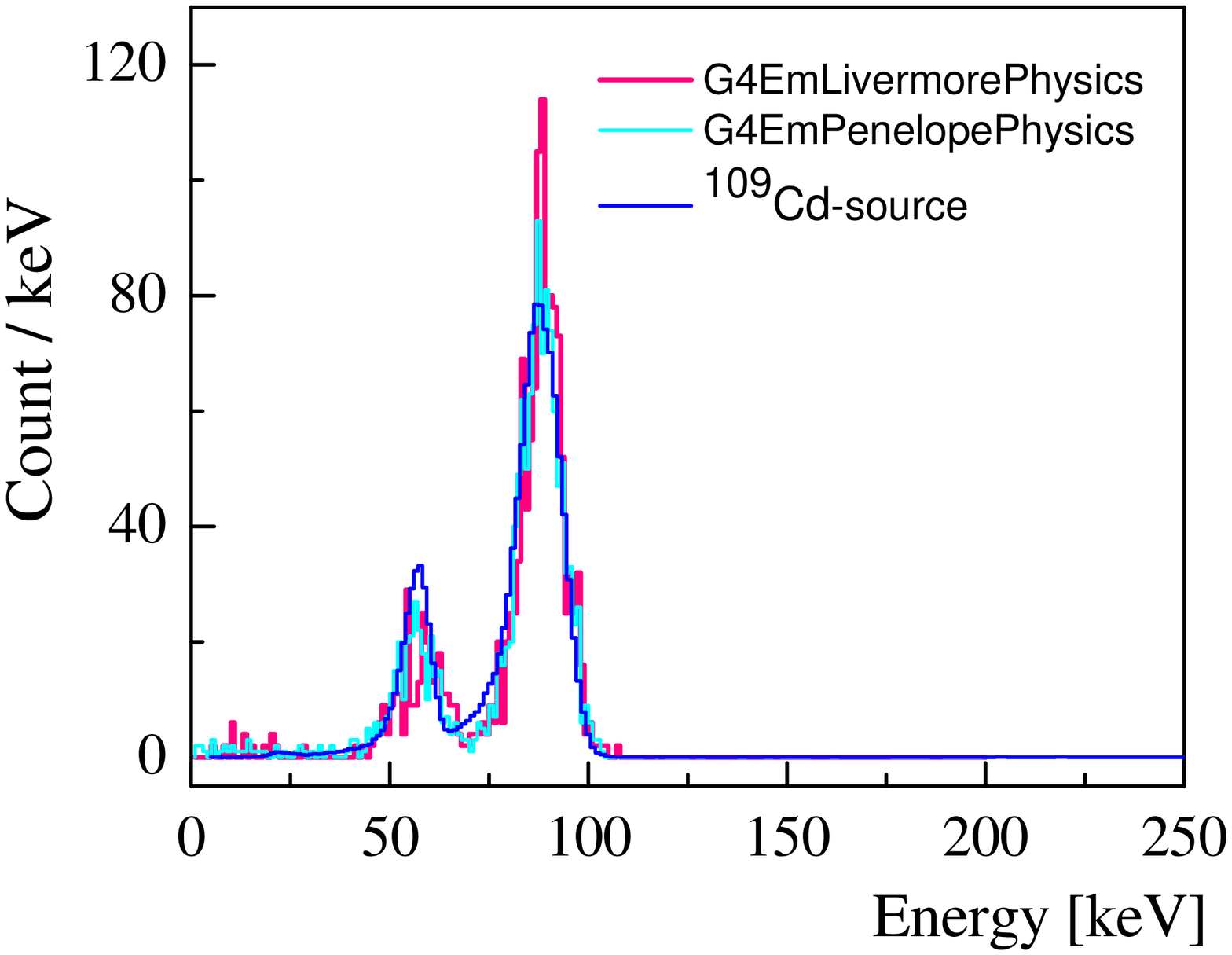} \vspace{-0.45pc}
\caption{\label{pic4} {\small
A comparison of experimental and simulated spectra of LPC obtained from $^{109}$Cd-source.
Turquoise and red histograms are the Geant4 simulations obtained with the Penelope and Livermore model respectively. Blue histogram is experimental data from $^{109}$Cd-source.}
}
\vspace{-0.0pc}
\includegraphics[width=4.5pc,angle=270]{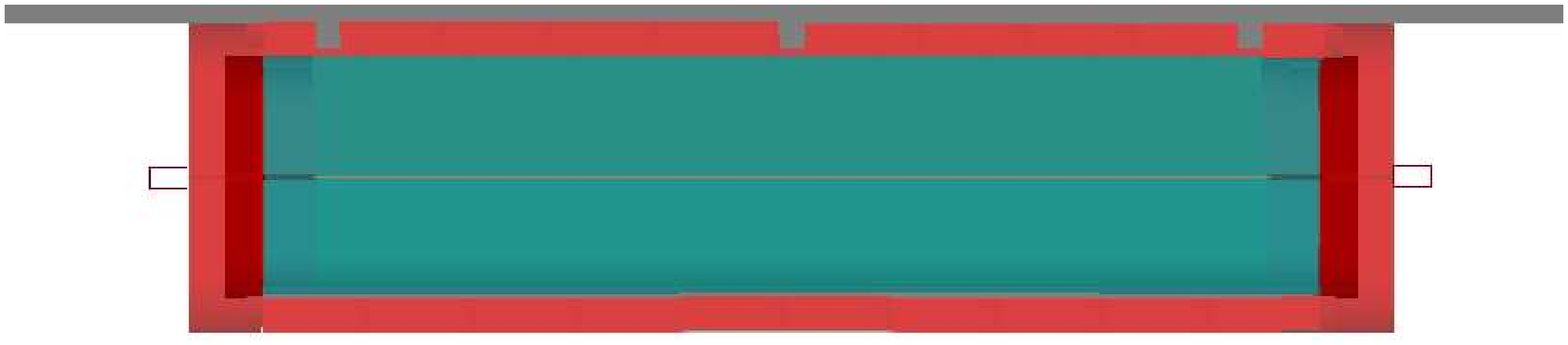}
\vspace{-0.2pc}
\caption{\label{pic3} {\small Calibration channel and calibration holes.}
}
\end{wrapfigure}
simulate the interaction of neutrons with matter, this library was chosen based on \cite{r4}.
The radioactive source (nuclear decay, alpha, beta, and gamma radiation) is specified in a mac-file using the General Particle Source library.
The models we created on the basis of two libraries allowed us to obtain a good agreement between the calculated and experimental response functions from an external calibration gamma source in most of the analyzed cases.

Figure \ref{pic4} shows a comparison of the measured and simulated spectra from an external $^{109}$Cd-source.
The source is placed above the detector in the calibration channel, which is located inside the low-background shield, Fig.\ref{pic3}. To reduce the absorption of gamma-quanta in the shield substance, three calibration holes are made above the center of the detector and along the edges of the sensitive zone of the detector. In standard mode, calibration is carried out in the center hole. The same was done in the model.

As noted in the introduction, the purpose of our experiment
is the search for rare events from $2\nu2K$ capture using registration a time correlation (coincidence) between the two $K$ fluorescence quantum that
is emitted after the decay of double $K$ vacancy states and Auger electrons.
The magnitude of the probable effect is estimated on the basis of the results of comparative analysis of background data of the proportional counter
filled with samples of pure xenon with different concentrations of the studied isotope. The ultimate sensitivity of the setup to the effect of interest depends primarily on the intrinsic detector background and the
quality of the methods used to separate the desired signal with a specified set of characteristics.
This requires an extra small gamma-background.
In the same time, a double $K$-shell photoionization of the
atom can create the ``hollow atom'' by absorbing a single
photon and releasing both $K$ electrons.
Relaxation of the excited electron shell of a daughter atom can mimic the useful signal.

\begin{wrapfigure}[19]{l}{0.5\textwidth} \vspace{-0.0pc}
\vspace{-0.2pc}
\centering
\includegraphics[width=15pc,angle=0]{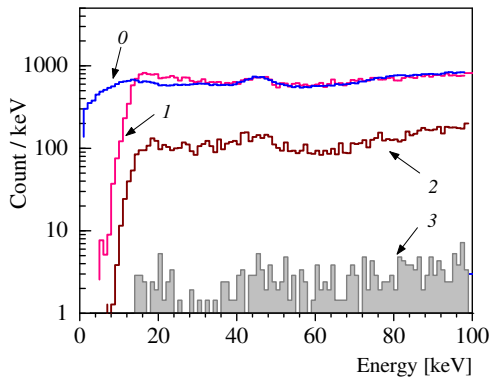} \vspace{-0.3pc}
\caption{\label{pic5} {\small Spectrum of the background
of LPC filled with xenon: (\emph{1}) - all events, (\emph{2}) - two-point events, (\emph{3}) - three-point
events, (\emph{0}) - Monte-Carlo simulations spectrum of  low-energy gamma-quantum for interactions in the
detector.}
}
\vspace{-0.0pc}
\end{wrapfigure}

\textbf{Gamma-background simulation.}\\
Gamma ray background from  the structural elements of the shell of the detector as well as in the elements of the shield (copper, lead) is almost exclusively coming from the uranium and thorium decay series as well as from decay of $^{40}$K.
External gamma backgrounds originate from these radionuclides in the surrounding rock of deep underground laboratories.
Therefore, our purpose was to estimate the response of the LPC to the gamma background from surrounding material.
The activity of natural radioactive isotopes content in different source materials of the low-background shield  was measured by the ultra-low background germanium gamma-spectrometer \cite{HPGe}.
The simulation included the calculation of decays of members of uranium-radium and actinium natural radioactivity
chains contained in the structural elements of the detector shell as well as in the elements of the shield.
Figure \ref{pic5} shows the background
spectrum accumulated for one year of measurement (curve \emph{1}) using LPC and Monte Carlo simulation of gamma-spectrum (curve \emph{0}).
As can be seen in the figure, the simulated and experimental spectrum of all events with energies above 30 keV coincides quite well.
The difference between the spectra at low energies can be related to the fact that at the ends of the detector the proportional amplification  changes to the ionization mode, which was not taken into account in this simulation. Also, edge effects at the ends were not considered.

The same figure shows the spectra of two- and three-point events, curve \emph{2} and \emph{3} respectively.
Events that are associated with a photon absorption
are distributed among the groups of one- and two-point events
in accordance with the probabilities of the Compton process,
photoabsorption, and the probability for the characteristic
radiation yield upon the photoabsorption induced filling of
the $K$ vacancy in the target atom.

Two-point events take place if, for example, a photon
undergoes the Compton scattering and it is absorbed in the
counter's volume, a photon undergoes photoeffect and
fluorescence quantum are emitted in the process of the vacancy filling or
an electron creates the bremsstrahlung. In all these cases,
a secondary photon is absorbed in some distance from its origin
and creates the second cluster of ionization. If these two clusters
are located at a different distance from the anode, the detected
signal will be made of two pulses, with the time difference
corresponding to the moments of entering the gas amplification
region by two groups of primary electrons.

Three-point events can appear if all secondary radiation is
absorbed inside the counter, as in, Compton scattering of
a photon, followed by the photoeffect and release of the fluorescence quantum by filling the vacancy in the atomic shell, photoeffect followed by the \emph{K}-shell ionization, by the photoelectron in
another atom and release of two fluorescence quantum. Three-point events,
which are the main goal of our study, have to be formed as a
result of absorption into the fiducial volume of the counter of
two characteristic photons and the Auger electrons generated
by the double \emph{K}-shell vacancy production in the investigated xenon samples.

All more-than-three-point events are considered as multipoint
events. They can appear, for example, as a result of
Compton scattering of the photon by the \emph{K} electron, followed by the \emph{K} shell photoeffect and release of two \emph{K} fluorescence quantum. Events can move around within multipoint groups because of possible overlaps of the pulses of separate components of multipoint events. A detailed description of the selection of multipoint  events and their components  can be found, for example, in \cite{PTE}.

\textbf{Neutron background simulation.}\\
The $^{125}$I may be one of the most critical background sources in the energy region of interest for the search the $2\nu2K$ capture in $^{124}$Xe. Since the energy and type of emission of its decay products are similar to the expected effect of $2\nu2K$ capture.
In our detector, this isotope may form after the decay of $^{125}$Xe.
It is formed when a thermal neutron is captured by the $^{124}$Xe  with a sufficiently large probability ($\sigma _{th} = 165 \pm 11$ bn \cite{Atlas}).
This can make to the impossibility to separate the desired signal from $2K$ capture with a significant flux of background thermal neutrons.
The half-life of $^{125}$Xe is about 17 h, after which it decays into an atom $^{125}$I.
Its half-life is 59.49 days and it decays by electron capture (100\%)
to the first excited state of the
daughter $^{125}$Te which then emits a 35.5 keV gamma ray photon in dropping to its ground state in $1.49\times10^{-9}$ seconds.
Simultaneous registration of the 35.46 keV gamma-line, characteristic $K$ fluorescence quanta, and Auger electrons in working volume of LPC can generate a three-point event with a signature that simulates an event from a $2\nu2K$ capture of $^{124}$Xe.

Therefore, we decided to test the possible number of neutrons which can ``penetrate'' into the working volume of the detector, thermalize and interact with the $^{124}$Xe.
Several possible sources of neutrons were considered in the simulation calculations.
The first of them is neutron production in $(\alpha-n)$-reactions during the decay of the U/Th series and their daughter nuclei (which undergo $\alpha$-decay) in the detector case, as well as in the shield elements. These calculations showed that we have zero effect.
The next stage included the calculation of the penetration of neutrons produced in the walls of the low-background underground laboratory and evenly distributed in the air.

In the calculations, we focused on the work in which measurements were made of neutrons of different energy ranges, in the same laboratory where the installation for searching for $2K$ capture in $^{124}$Xe is located.
In the first work, the measurements of the thermal neutron flux were carried out on the neutron background measurements with the [ZnS(Ag)+$^6$LiF] scintillation detector \cite{r6}.
As a result of the measurements, an estimate was obtained for the thermal neutron flux at the level $(2.6\pm0.4)\times10^{-5}$ cm$^{-2}$s$^{-1}$.
In the second work \cite{r7}, measurements of neutrons with energies of $\sim700$ keV were carried out. The value of the neutron flux with such an energy of $5.3\times10^{-7}$ - $1.8\times10^{-7}$ cm$^{-2}$s$^{-1}$ was obtained.

These values were incorporated into the design model of our installation. The neutron layer surrounding the shield from all sides was selected with a thickness of 100 mm.
Monte Carlo simulation of spectra of neutron passed through the material of passive  shield and scattered in the LPC allowed us to obtain the following results. The thermal neutrons (in reality neutrons with energies below 0.5 eV were taken) do not penetrate through the low-background shield into the detector, and therefore do not contribute to the production of the $^{125}$Xe.
Neutrons with an energy of $\sim700$ keV can penetrate into the detector through the low-background shield. We have $\sim2.5$ thermalized neutrons in the detector when normalizing the simulation results for a year of measurements.
We will have 0.02 atoms of the $^{125}$I in the detector in one year of measurements, taking into account the interaction cross-section and the number of $^{124}$Xe atoms.
This will give us no more than one atom of the $^{125}$I in our detector in 4-5 years of measurements.

\section{Conclusions}

The reducing of the background is crucial when searching for the  $2\nu2K$ capture in $^{124}$Xe because it determines the sensitivity when the total amount of $^{124}$Xe is equal to fixed.
To estimate the radioactive contamination of the experimental setup ``2K-CAPTURE''  has been applied the Geant4-based Monte Carlo simulation of the energy spectra from background sources.
The energy distributions of the detector response obtained in the Geant4 MC simulation were compared with measurements from a calibration source, which resulted in a good description of the data. Detector response functions of thermal neutrons penetrating into the working volume of the LPC  have also been calculated.


\end{document}